\documentclass[letterpaper,english,reprint, aps]{revtex4-2}
\pdfoutput=1
\usepackage[T1]{fontenc}
\usepackage[latin9]{inputenc}
\usepackage{amsmath}
\usepackage{amssymb}
\usepackage{graphicx}

\makeatletter

\pdfpageheight\paperheight
\pdfpagewidth\paperwidth

\makeatother

\usepackage{babel}
\begin{document}
\title{Absorption and optical selection rules of tunable excitons in biased
bilayer graphene}
\author{J. C. G. Henriques$^{1}$, Itai Epstein$^{2,3,4}$ and N. M. R. Peres$^{1,5}$}
\affiliation{$^{1}$Department and Centre of Physics, University of Minho, Campus
of Gualtar, 4710-057, Braga, Portugal}
\affiliation{$^{2}$School of Electrical Engineering, Faculty of Engineering, Tel
Aviv University, Tel Aviv 6997801, Israel}
\affiliation{$^{3}$Center for Light-Matter Interaction, Tel Aviv University, Tel
Aviv 6997801, Israel}
\affiliation{$^{4}$QuanTAU, Quantum Science and Technology center, Tel Aviv University,
Tel Aviv 6997801, Israel}
\affiliation{$^{5}$International Iberian Nanotechnology Laboratory (INL), Av. Mestre
Jose Veiga, 4715-330, Braga, Portugal}
\begin{abstract}
Biased bilayer graphene, with its easily tunable band gap, presents
itself as the ideal system to explore the excitonic effect in graphene
based systems. In this paper we study the excitonic optical response
of such a system by combining a tight binding model with the solution
of the Bethe-Salpeter equation, the latter being solved in a semi-analytical
manner, requiring a single numerical quadrature, thus allowing for
a transparent calculation. With our approach we start by analytically
obtaining the optical selection rules, followed by the computation
of the absorption spectrum for the case of a biased bilayer encapsulated
in hexagonal boron nitride, a system which has been the subject of
a recent experimental study. An excellent agreement is seen when we
compare our theoretical prediction with the experimental data.
\end{abstract}
\maketitle

\section{Introduction}

Since graphene was first isolated \citep{novoselov2004electric} many
other two dimensional materials have be discovered. Examples of these
are the semiconductor monolayer transition metal dichalcogenides (TMDs)
\citep{Wang2018Colloquium}, the highly anisotropic phosphorene \citep{carvalho2016phosphorene},
or the insulator hexagonal boron nitride (hBN) \citep{caldwell2019photonics}.
Despite their own unique properties, these three materials have something
in common, all of them have an optical response dominated by excitonic
effects.

In the simplest possible picture, an exciton is formed upon the excitation
of an electron to the conduction band, leaving a hole in the valence
band. These two particles, which have opposite charges, interact via
an electrostatic potential \citep{Cudazzo2011}. This situation is
somewhat similar to what is found in the Hydrogen atom, and leads
to the formation of bound states inside the band gap of the material.
Due to the reduced screening in the out-of-plane direction, two dimensional
materials host excitons with large binding energies , easily surpassing
the energies usually found in their three dimensional counterparts.
Furthermore, excitons in these 2D systems couple efficiently with
light resulting in huge oscillator strengths. This has made the exploration
of excitonic effects a rich research field, with many possible applications,
such as deep-UV optoelectronics \citep{watanabe2004direct,kubota2007deep,caldwell2019photonics}
in hBN, or the exploration of valleytronics \citep{schaibley2016valleytronics}
and single photon emitters \citep{yu2017moire,branny2017deterministic}
in TMDs, which are of great importance in the field of quantum information.
Furthermore, it has been recently shown that the combination of a
TMD with a Van der Waals heterostructure cavity enhances the light-mater
interaction, thus allowing for unitary excitonic absorption \citep{epstein2020near}.

Although present in many 2D materials, low energy excitonic bound
states are noticeably absent in pristine monolayer graphene, something
easily understood if one recalls that this material is a semi-metal,
thus lacking the necessary band gap for the formation of such states.
This, however, does not stop us from studying excitonic physics in
graphene based systems, since through material engineering many systems
can be designed to obtain the desired properties. A clear example
of this is the case of biased bilayer graphene (BBLG), where two graphene
monolayers are stacked and a displacement field is applied to them.
Even though bilayer graphene (BLG) inherits the lack of a band gap
from its monolayer form, the presence of the displacement field opens
a gap in the system, which can be finely tuned from the mid to far-infrared
\citep{castro2007biased,zhang2009direct,oostinga2008gate}. This finite
gap is the key ingredient to combine the desirable properties of excitons
with the unique physics of graphene based systems.

The first theoretical study about excitons in BBLG was carried out
in Ref. \citep{park2010tunable}, where a pronounced resonance, sensitive to the displacement field, was predicted to appear inside
the band gap in absorption measurements. More recently, in Ref. \citep{ju2017tunable},
an experimental study was carried out to characterize the excitonic
optical response of BBLG encapsulated in hBN. Our goal with this paper
is to produce a theoretical description of the experimental data measured
in that work. To achieve this, we will describe the electronic properties
using a tight binding model, while solving the Bethe-Salpeter equation
(BSE) to obtain the energies and wave functions of the excitons. While
oftentimes the BSE is coupled to DFT calculations , and solved in
a fully numerical manner \citep{park2010tunable,Fuchs2013,Komsa2013,Alejandro2016,di2020optical},
requiring huge computational effort, our approach uses a simple analytical
treatment which requires a single numerical quadrature, thus greatly
reducing the numerical cost of the computation.

This paper is organized as follows. In Sec. \ref{sec:Tight-binding-model}
we introduce the tight binding model which describes the biased bilayer,
and diagonalize it. In Sec. \ref{sec:Bethe-Salpeter-equation}, we
start the discussion on the excitonic properties of the system, which
we obtain by solving the Bethe-Salpeter equation. In Sec. \ref{sec:Optical-Absorption},
we discuss the optical properties of the biased bilayer. First, we
discuss the optical selection rules; then we compute the absorption
spectrum and compare it with the experimental results of Ref. \citep{ju2017tunable}.
Finally, in Sec. \ref{sec:Conclusion}, we give an overview of the
paper and our closing remarks. An appendix describing how to efficiently
solve the Bethe-Salpeter equation closes the paper.

\section{Tight binding model\label{sec:Tight-binding-model}}

\begin{figure}[h]
\centering{}\includegraphics{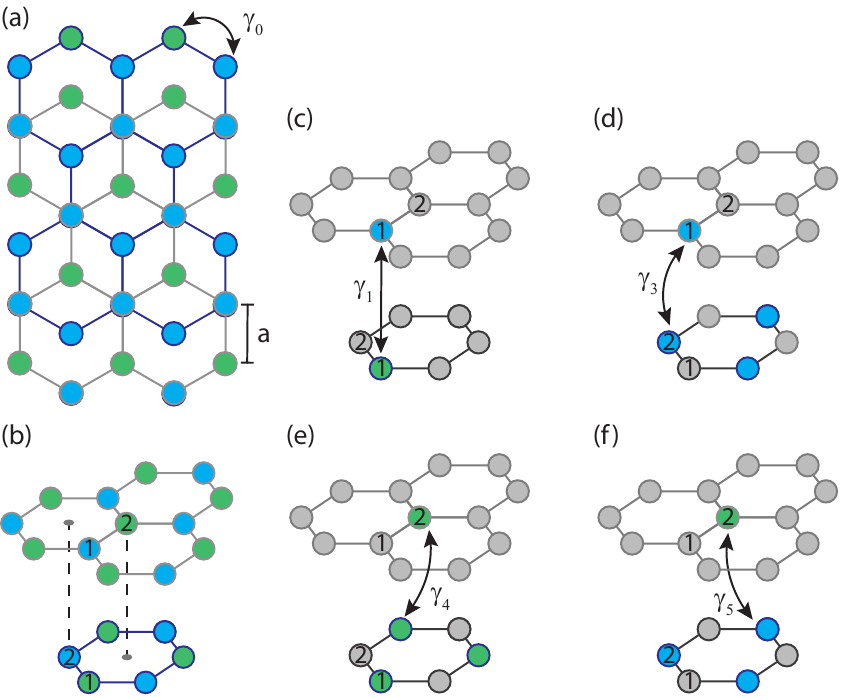}\caption{\label{fig:BBLG shcematic}(a) Top view of Bernal stacked bilayer
graphene. The atoms of different sub-lattices are colored in different
tones for clarity. Also depicted is the intralayer nearest neighbor
hopping $\gamma_{0}$. (b) Side view of the bilayer graphene. (c)-(f)
Schematic representation of nearest neighbor and next nearest neighbor
interlayer hoppings.}
\end{figure}

To start our discussion let us study the electronic properties of
a graphene bilayer subject to an external bias, which we schematically
depict in Figure \ref{fig:BBLG shcematic}. 
To characterize this system we shall employ a tight binding Hamiltonian
written directly in momentum space. To obtain this Hamiltonian we
account for nearest neighbors intralayer hoppings ($\gamma_{0}$),
as well as nearest neighbors $(\gamma_{1})$ and next nearest neighbors
($\gamma_{3}$, $\gamma_{4}$ and $\gamma_{5}$) interlayer hoppings.
Doing so, one finds the Hamiltonian to be:
\begin{equation}
H_{\textrm{TB}}=\left[\begin{array}{cccc}
V & \gamma_{0}\phi(\mathbf{k}) & \gamma_{1} & \gamma_{4}\phi^{*}(\mathbf{k})\\
\gamma_{0}\phi^{*}(\mathbf{k}) & V & \gamma_{3}\phi^{*}(\mathbf{k}) & \gamma_{5}\phi(\mathbf{k})\\
\gamma_{1} & \gamma_{3}\phi(\mathbf{k}) & -V & \gamma_{0}\phi^{*}(\mathbf{k})\\
\gamma_{4}\phi(\mathbf{k}) & \gamma_{5}\phi^{*}(\mathbf{k}) & \gamma_{0}\phi(\mathbf{k}) & -V
\end{array}\right],
\end{equation}
when written in the basis $\{|1,b\rangle,|2,b\rangle,|1,t\rangle,|2,t\rangle\}$
(with $b$ and $t$ denoting the bottom and top layers, respectively,
and the labels $1$ and $2$ are defined as in Fig. \ref{fig:BBLG shcematic}).
Here, $V$ quantifies the bias and $\phi(\mathbf{k})=e^{iak_{y}}+e^{ia(k_{x}\sqrt{3}-k_{y})/2}+e^{-ia(k_{x}\sqrt{3}+k_{y})/2}$
is a phase factor determined by the honeycomb geometry of the individual
layers. Since we will be mostly interested in the low energy response
of the system, we shall restrict our study to the vicinity of the
Dirac points of the the first Brillouin zone. Mathematically, this
amounts to approximating the phase factor $\phi(\mathbf{k})$ as $\phi(\mathbf{k})\approx\frac{3}{2}a(\tau k_{x}+ik_{y})$,
with $\tau=\pm1$ labeling the two Dirac points. Note that henceforth,
the momentum $\mathbf{k}=(k_{x},k_{y})$ is measured relatively to
the chosen Dirac point. With this approximation one finds the low
energy Hamiltonian
\begin{equation}
\frac{H_{\textrm{low}}}{\hbar v_{F}}=\left[\begin{array}{cccc}
\frac{V}{\hbar v_{F}} & \tau ke^{\tau i\theta} & \frac{\gamma_{1}}{\hbar v_{F}} & \frac{\gamma_{4}}{\gamma_{0}}\tau ke^{-\tau i\theta}\\
\tau ke^{-\tau i\theta} & \frac{V}{\hbar v_{F}} & \frac{\gamma_{3}}{\gamma_{0}}\tau ke^{-\tau i\theta} & \frac{\gamma_{5}}{\gamma_{0}}\tau ke^{\tau i\theta}\\
\frac{\gamma_{1}}{\hbar v_{F}} & \frac{\gamma_{3}}{\gamma_{0}}\tau ke^{\tau i\theta} & -\frac{V}{\hbar v_{F}} & \tau ke^{-\tau i\theta}\\
\frac{\gamma_{4}}{\gamma_{0}}\tau ke^{\tau i\theta} & \frac{\gamma_{5}}{\gamma_{0}}\tau ke^{-\tau i\theta} & \tau ke^{\tau i\theta} & -\frac{V}{\hbar v_{F}}
\end{array}\right],\label{eq:low energy Hamiltonian}
\end{equation}
where we introduced the Fermi velocity defined as $\hbar v_{F}=3a\gamma_{0}/2$,
and $\theta=\arctan k_{y}/k_{x}$. For simplicity, in what follows,
we will be mainly concerned with the case where $\gamma_{3}=\gamma_{4}=\gamma_{5}=0$,
i.e. accounting only for nearest neighbors hoppings. However, later
in the text, we shall see how these hopping parameters affect the
response of the system.

\begin{figure}[h]
\centering{}\includegraphics{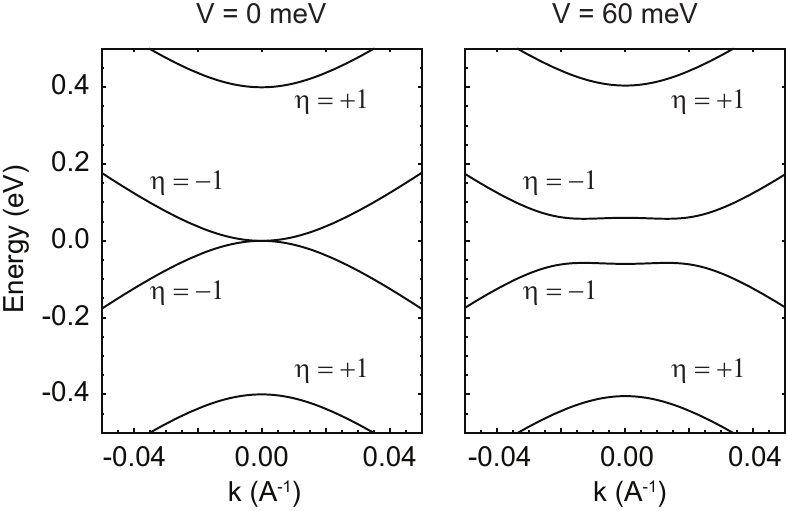}\caption{\label{fig:BBLG_bands}Electronic bands near the Dirac valley associated
with $\tau=1$. On the left panel we show the case without bias, and
on the right one the bands when a bias of $V=60$ meV is considered.
For convenience we label the valence and conduction bands with an
additional index $\eta$, which is equal to $-1$ for the bands closer
to the middle of the gap and $1$ for the other two.}
\end{figure}

Diagonalizing the low energy Hamiltonian of Eq. (\ref{eq:low energy Hamiltonian}),
we find the electronic bands depicted in Fig. \ref{fig:BBLG_bands},
which were obtained using $\gamma_{0}=3$ eV and $\gamma_{1}=0.4$
eV. 
In this figure, we see that, for zero bias, one finds the usual bands
of bilayer graphene, which host no gap since the lowest conduction
band and the highest valence band touch at $k=0$. When the bias is
increased to $V=60$ meV, the two aforementioned bands are deformed,
taking the shape of a ``Mexican hat''. This deformation of the valence
and conduction bands leads to the opening of a gap, something crucial
to exploit excitonic effects. If the next nearest neighbors hoppings
had been considered only minor changes would appear in the band structure.
Even though small, these modifications in the bands have an important
effect on the optical selection rules (which we discuss ahead), since
they introduce the trigonal warping effect, which reduces the symmetry
of our calculation from $C_{\infty}$ to $C_{3}$.

From the diagonalization of Eq. (\ref{eq:low energy Hamiltonian}),
we also find the eigenvectors associated with each band. We label
these vectors as $|c,\eta,\mathbf{k\rangle}$ and $|v,\eta,\mathbf{k}\rangle$,
where the index $\eta$ is defined as in Fig. \ref{fig:BBLG_bands},
i.e. the bands closer to the middle of the gap have $\eta=-1$, while
the other two have $\eta=1$; they read:
\begin{align}
|c,-1,\mathbf{k}\rangle & =\left[a_{c,1}^{-}e^{i\tau\theta},a_{c,2}^{-},a_{c,3}^{-}e^{i\tau\theta},a_{c,4}^{-}e^{2i\tau\theta}\right]\label{eq:low_spinors}\\
|c,+1,\mathbf{k}\rangle & =\left[a_{c,1}^{+},a_{c,2}^{+}e^{-i\tau\theta},a_{c,3}^{+},a_{c,4}^{+}e^{i\tau\theta}\right]\\
|v,-1,\mathbf{k}\rangle & =\left[a_{v,1}^{-}e^{-i\tau\theta},a_{v,2}^{-}e^{-2i\tau\theta},a_{v,3}^{-}e^{-i\tau\theta},a_{v,4}^{-}\right]\\
|v,+1,\mathbf{k}\rangle & =\left[a_{v,1}^{+},a_{v,2}^{+}e^{-i\tau\theta},a_{v,3}^{+},a_{v,4}^{+}e^{i\tau\theta}\right]
\end{align}
The specific definition of each entry is not of particular interest
to the current analysis, however, the phase of each spinor was chosen
carefully. As a consequence of the definition of $\theta$, as $k\rightarrow0$
the complex exponential $e^{i\tau\theta}$ is discontinuous. To avoid
this discontinuity in the eigenvectors, the phase of each one was
chosen such that the complex exponentials appear multiplied by terms
that vanish in the limit of small $k$. It is by now known that the
phase choice of the Bloch factors plays a crucial role in determining
the optical selection rules, and the criteria we established here
is the one that directly leads to Hydrogen-like selection rules in
the monolayer. Although other phase choices can be employed, they
require the introduction of an additional angular quantum number associated
with the pseudospin texture of the excitonic states \citep{park2010tunable,Zhang2018,Cao2018}.
Even though we do not shown the explicit form of the entries of each
eigenvector, in light of the discussion regarding where the complex
exponentials are place, it should be clear that $|a_{c,1}^{-}|,|a_{c,3}^{-}|,|a_{c,4}^{-}|\ll|a_{c,2}^{-}|$,
$|a_{c,2}^{+}||a_{c,4}^{+}|\ll|a_{c,1}^{+}|,|a_{c,3}^{+}|$, $|a_{v,1}^{-}|,|a_{v,2}^{-}|,|a_{v,3}^{-}|\ll|a_{v,4}^{-}|$
and $|a_{v,2}^{+}||a_{v,4}^{+}|\ll|a_{v,1}^{+}|,|a_{v,3}^{+}|$. These
relations between the amplitude of the different vector components
are enough for us to gain intuition about the coupling strength of
light with different excitonic states, which we introduce next.

\section{Bethe-Salpeter equation\label{sec:Bethe-Salpeter-equation}}

Now that the single particle regime was studied, we move on to the
excitonic part of the problem. To obtain the response of the system
due to excitons we must first determine their energies and wave functions.
These quantities can be obtained from the solution of the Bethe-Salpeter
equation. This integral equation in momentum space may be written
for a multi-band system as \citep{Pedersen2019,Thomas2015cond}:
\begin{align}
E\psi_{cv}(\mathbf{k}) & =(E_{c,\mathbf{k}}-E_{v,\mathbf{k}})\psi_{cv}(\mathbf{k})\nonumber \\
 & +\sum_{c',v'}\sum_{\mathbf{q}}V(\mathbf{k}-\mathbf{q})\langle c,\mathbf{k}|c',\mathbf{q}\rangle\langle v',\mathbf{q}|v,\mathbf{k}\rangle\psi_{cv}(\mathbf{q})\label{eq:BSE}
\end{align}
where, to simplify the notation, we omitted the index $\eta$ (which
is now included in the indexes $c$ and $v$). We note that this equation
takes the single particle energies $E_{c/v,\mathbf{k}}$ and eigenvectors
$|c/v,\mathbf{k}\rangle$ as the input, and couples the different
bands through an electrostatic potential $V(\mathbf{k}),$ thus capturing
the many body nature intrinsic to excitons. By solving this equation,
one finds the exciton energies, $E$, and the associated wave functions,
$\psi_{cv}(\mathbf{k})$. In our approach we consider the electrostatic
potential $V(\mathbf{k}-\mathbf{q})$ to be the Rytova-Keldysh potential
\citep{rytova1967,keldysh1979coulomb} (which is usually employed
to describe excitonic effects in monolayers \citep{Cudazzo2011}).
This potential can be obtained from the solution of the Poisson equation
for a charge embedded in a thin film, by taking the limit of vanishing
thickness. Although this potential presents a complex form in real
space, its representation in momentum space is rather simple
\begin{equation}
V(\mathbf{k})=\frac{\hbar c\alpha}{\epsilon}\frac{1}{k(1+r_{0}k)},
\end{equation}
where $\alpha\sim1/137$, $\epsilon$ is the mean dielectric constant
of the media above and below the bilayer and $r_{0}$ corresponds
to an in plane screening length which is related to the 2D polarizability
of the material; its value can be found from the single particle bands
and Bloch factors. The numerical value of $r_{0}$ is of great importance
to accurately describe the excitonic properties of a given system,
since it directly affects the screening experienced by the carriers.
Although the values of $r_{0}$ are well documented for TMDs, hBN
and other 2D materials, the same is not true for biased bilayer graphene,
where, in principle, $r_{0}$ varies with the bias $V$. In Ref. \citep{Li2019}
the value of $r_{0}$ as a function of $V$ was determined, however,
it was done using a 4 band tight binding model, which may not be sufficiently
accurate. Ideally the value of $r_{0}$ should be determined from
\emph{ab initio} calculations taking into account several valence
and conduction bands, because, as discussed in Ref. \citep{tian2019electronic},
depending on the material, a small number of bands of a minimal tight
binding model may or may not be enough to accurately determine $r_{0}$.
Since a detailed \emph{ab initio} study of $r_{0}$ in BBLG is lacking
in the literature, we will use the results of \citep{Li2019} as a
reference. There one finds that, as expected, $r_{0}$ decreases with
increasing band gap, or alternative, it decreases with increasing
bias in a non-trivial way.

To solve the BSE we start by noting that Eq. (\ref{eq:BSE}) corresponds
in fact to four equations, one for each possible pair $(v,c)$, forming
an eigenvalue problem. To more easily solve the BSE, we assume that
the excitons have a well defined angular quantum number $m$, such
that $\psi_{cv}(\mathbf{k})=f_{cv}(k)e^{im\theta}$. Using this, and
manipulating the phase choices of the spinors in Eq. (\ref{eq:low_spinors}),
one can show that the BSE can be reduced to a 1D integral equation,
which can then be efficiently solved with a single numerical quadrature
(the details on how to achieve this are discussed in the Appendix).

After solving the BSE for a wide range of values for the bias  (whilst using the results and parameters of the previous section,
as well as the $r_{0}$ of Ref. \citep{Li2019} and taking $\epsilon=6.9$
\citep{laturia2018dielectric}, corresponding to the case of hBN encapsulated
BLG), we observed that of the four possible sets of $\psi_{cv}(\mathbf{k})$,
the contribution of the one with $\eta_{c}=\eta_{v}=-1$ was by far
the dominant one, so much so, that the other wave functions appeared
to vanish in comparison. This is a reasonable result to find, since
intuition tells us that the bands with $\eta=-1$, i.e. the ones closer
to the middle of the gap, should dominate the low energy response.
Hence, to further optimize the calculation, one may restrict the BSE
to a single pair of bands, thus greatly reducing the computational
cost of the problem, without hindering the quality of the results.

\section{Optical Absorption\label{sec:Optical-Absorption}}

In this section our goal is to determine the optical selection rules
of the biased bilayer and to compute its optical absorption. To achieve
both these goals we shall start by looking at the optical conductivity
of the system. In the dipole approximation, and considering normal
incidence, the optical conductivity follows as \citep{Thomas2015cond}
\begin{equation}
\sigma(\hbar\omega)\propto\sum_{n}\frac{\boldsymbol{\Omega}_{n}\boldsymbol{\Omega}_{n}^{*}}{\hbar\omega-E_{n}+i\frac{\Gamma_{n}}{2}}+\left(\omega\rightarrow-\omega\right)^{*},
\end{equation}
where the sum over $n$ accounts for the different excitonic contributions
with energy $E_{n}$, and the second term corresponds to the non-resonant part of the optical response. The quantity $\boldsymbol{\Omega}_{n}$ is defined as
\begin{equation}
\boldsymbol{\Omega}_{n}=\sum_{c,v}\sum_{\mathbf{k}}\psi_{cv}(\mathbf{k})\langle v,\mathbf{k}|\mathbf{r}|c,\mathbf{k}\rangle,\label{eq:Omega_n}
\end{equation}
where $\langle v,\mathbf{k}|\mathbf{r}|c,\mathbf{k}\rangle$ is the
interband matrix element of the dipole operator. Also, we introduced
a phenomenological linewidth, $\Gamma_{n}$, which we assume to be
$n-$dependent, to more easily compare our theoretical predictions
with experimental data. As we have discussed before, we consider only
the contribution of the bands with $\eta=-1$, since the $\psi_{cv}$
associated with the other ones are essentially zero. To evaluate the
dipole matrix element we use the relation
\begin{equation}
\langle v,\mathbf{k}|\mathbf{r}|c,\mathbf{k}\rangle=\frac{\langle v,\mathbf{k}|\left[H,\mathbf{r}\right]|c,\mathbf{k}\rangle}{E_{v,\mathbf{k}}-E_{c,\mathbf{k}}}.\label{eq:Dipole MatEl}
\end{equation}
Although when solving the BSE we considered only the contribution
of the nearest neighbor hoppings $\gamma_{0}$ and $\gamma_{1}$,
to understand the role played by the other hoppings in the optical
response, we shall take them into account when the commutator $\left[H,\mathbf{r}\right]$
is evaluated.

\begin{figure}[h]
\centering{}\includegraphics{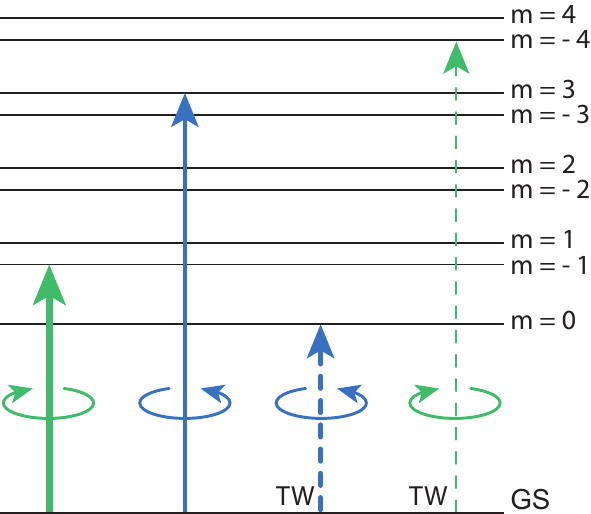}\caption{\label{fig:Selection rules}Schematic representation of the optical
selection rules in the Dirac valley with $\tau=1$. Dashed lines indicate
processes which are only possible with trigonal warping (TW). The thickness
of the arrows illustrates the strength of the coupling with photons.
The results for the $\tau=-1$ valley follow from these ones through
the substitution $m\rightarrow-m$, and by switching the two circular
polarizations. The ground state, i.e. the excitonic vacuum, is represented by GS}
\end{figure}

Let us now consider that our system is excited by a circularly polarized
electric field. The relevant dipole matrix elements for such a situation
are $\langle v,\mathbf{k}|x\pm iy|c,\mathbf{k}\rangle$, corresponding
to positive and negative circular polarization. Using Eq. (\ref{eq:Dipole MatEl})
to evaluate the matrix element, we find that the terms proportional
to $\gamma_{0}$ give the contribution
\begin{align}
\langle v,\mathbf{k}|x\pm iy|c,\mathbf{k}\rangle\big|_{\gamma_{0}} & \propto\left(a_{c,2}^{-}a_{v,1}^{-}+a_{c,3}^{-}a_{v,4}^{-}\right)e^{i\tau\theta}\left(\tau\mp1\right)\nonumber \\
 & +\left(a_{c,1}^{-}a_{v,2}^{-}+a_{c,4}^{-}a_{v,3}^{-}\right)e^{3i\tau\theta}\left(\tau\pm1\right).
\end{align}
It is then clear from Eq. (\ref{eq:Omega_n}), that such a result
implies that only states with $m=-\tau$ and $m=-3\tau$, i.e. $p-$
and $f-$states, can be excited. From this result, one also sees that
positive polarization excites states with $m=1$ in the valley with
$\tau=-1$, and states with $m=-3$ in the opposite valley; an analogous
results holds negative polarization. Regarding the coupling strength
between the photons and the allowed exciton states, and recalling
the comments made on the magnitude of eigenvector entries after Eq.
(\ref{eq:low_spinors}), one can expect the resonances associated
with the $p-$states to be far more intense than the ones associated
with the $f-$states, which should present a vanishingly small oscillator
strength. The contributions to the matrix element $\langle v,\mathbf{k}|x\pm iy|c,\mathbf{k}\rangle$
proportional to $\gamma_{3}$ and $\gamma_{4}$, lead to the same
selection rules as the ones we have just discussed. Furthermore, since
$\gamma_{1}\gg\gamma_{3},\gamma_{4}$, one finds that these hoppings
can be safely ignored when studying the linear optical response of
the bilayer. In stark contrast with this observation, the part of
the dipole matrix element which is proportional to $\gamma_{5}$ introduces
new selection rules. Explicitly:
\begin{align}
\langle v,\mathbf{k}|x\pm iy|c,\mathbf{k}\rangle\big|_{\gamma_{5}} & \propto a_{c,2}^{-}a_{v,4}^{-}\left(\tau\pm1\right)\nonumber \\
 & +a_{c,4}^{.-}a_{v,2}^{-}e^{4i\tau\theta}\left(\tau\mp1\right),
\end{align}
where we see that by accounting for the contribution of $\gamma_{5}$,
the excitation of $s-$ ($m=0$) and $g-$states $(m=\pm4)$ becomes
possible. Even though these contributions are proportional to $\gamma_{5}/\gamma_{0}\sim0.1$,
the large magnitude of $a_{c,2}^{-}a_{v,4}^{-}$ makes the $s-$states
rather relevant for the optical response; the $g-$states, however,
go totally unnoticed. A summary of the optical selection rules is
depicted in Fig. \ref{fig:Selection rules} for $\tau=1$. 
Since the two valleys are related through time reversal symmetry,
the results for the $\tau=-1$ valley are identical given that one
changes $m$ by $-m$, and switches the roles of the two circular
polarizations. Moreover, since linear polarization can be viewed as
the linear combination of the two circular polarization components,
the analysis we performed is easily applied to the case of linearly
polarized electric fields.

Having analyzed the optical selection rules of the biased bilayer,
and making use of the solutions of the BSE, the optical conductivity
is readily obtained. A quantity which is determined by the optical
conductivity and is experimentally more relevant is the optical absorption.
Here, we define the absorption as $\mathcal{A}=1-|r|^{2}-|t|^{2}$,
where $r$ and $t$ are the reflection and transmission coefficients,
respectively. These two coefficients are found from the longitudinal
conductivity as \citep{gonccalves2016introduction}:
\begin{align}
r & (\hbar\omega)=\frac{\alpha\pi\sigma(\hbar\omega)/\sigma_{0}}{2\sqrt{\epsilon}+\alpha\pi\sigma_{L}/\sigma_{0}}\\
t(\hbar\omega) & =\frac{2\sqrt{\epsilon}}{2\sqrt{\epsilon}+\alpha\pi\sigma(\hbar\omega)/\sigma_{0}},
\end{align}
with $\alpha\sim1/137$, $\sigma(\hbar\omega)$ the optical conductivity
and $\sigma_{0}=e^{2}/4\hbar$ the conductivity of a graphene monolayer.

\begin{figure}[h]
\centering{}\includegraphics{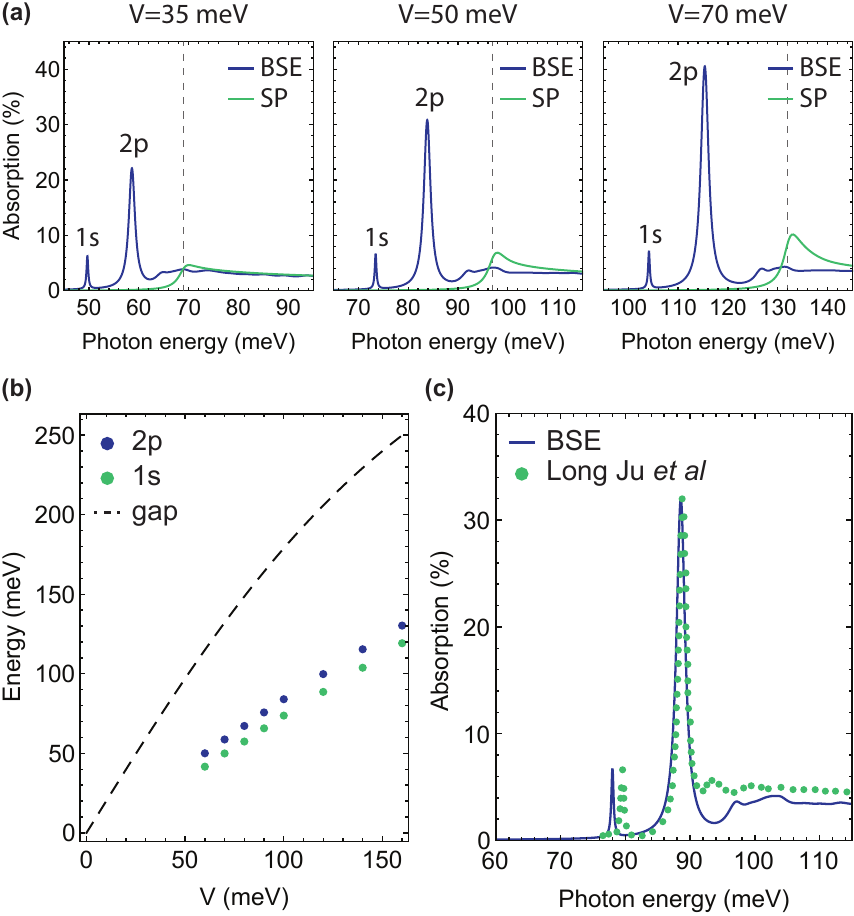}\caption{\label{fig:absorption}(a) Optical absorption of hBN encapsulated
BBLG for three different values of the bias $V$. The blue line corresponds
to the excitonic response obtained by using the solutions of the BSE;
the green line gives the optical absorption in the single particle
(SP) regime, obtained from Fermi's golden rule. The dashed line indicates
the location of the band gap. (b) Location of the 1s and 2p resonances
as a function of the bias $V$. The dashed line gives the magnitude
of the band gap. (c) Comparison between the experimental data of Ref.
\citep{ju2017tunable}, and our theoretical prediction for a bias
of $V=52$ meV.}
\end{figure}

Considering an in-plane linearly polarized electric field (for example
along the $x$ direction), using once again $\epsilon=6.9$ (which
describes a system encapsulated in hBN), and considering $\gamma_{5}=0.1\gamma_{1}$ \citep{Li2019},
we find the absorption spectra depicted in Fig. \ref{fig:absorption}
(a).
Analyzing these results, we see that for each bias two resonances
are well resolved, the smaller one at lower energies being due to
the $1s$ exciton, and the larger one corresponding to the excitation
of the $2p$ states. The linewidths of these resonances were chosen
in order to match the experimentally measured values of Ref .\citep{ju2017tunable},
which read $\Gamma_{1s}=0.4$meV and $\Gamma_{2p}=1.3$meV. Although
optically allowed, the $f-$ and $g-$states do not originate any
noticeable resonance due to their small oscillator strength; the same
is true for the more excited $s-$states. The more energetic $p-$states
present a significant oscillator strength, however, since these are
too close together, and too close to the band gap, they can hardly
be resolved; instead, these states form the beginning of a plateau.
This structure with an approximately constant magnitude is then extended
beyond the band gap, due to the contribution of the interband excitations.
Superimposed on the excitonic response, we also depict the interband
conductivity obtained in the single particle (SP) picture using Fermi's
golden rule. This allows us to see that the excitonic effect is not
only responsible for the resonances inside the gap, but also leads
to a different response at higher energies, as a result of the shift
in oscillator strength from the interband response to the excitonic
resonances. The results of Fig. \ref{fig:absorption} (a) agree well with those of \citep{park2010tunable}, which were obtained through more complex numerical approaches. Those results, however, do not account for the s-resonance which we include in our calculation.

Focusing now on the dependence of the absorption with the the bias,
we observe that as the bias is increased the main features of the
absorption remain qualitatively the same, but the optical response
is shifted to higher energies. Further evidence of this behavior is
given in panel (b) of the same figure where we depict the position
of the $1s$ and $2s$ resonances as a function of the bias. This
allows us to confirm that the resonances are in fact shifted to higher
energies as the bias increases, while also becoming more separated
from each other. The shift of the resonances to higher energies with
increasing bias was to be expected, since, as we also depict, the
band gap grows as the bias is turned up. However, we must note that
the trend of the peak's position differs slightly from that of the
band gap. This is effect is due to dependence of $r_{0}$ with $V$,
since as the bias increases, $r_{0}$ decreases, leading to more tightly
bound excitons, appearing deeper into the gap. These two competing
effects result in the almost linear dependence of the location of
the resonances with $V$.

Finally, in panel (c) of Fig. \ref{fig:absorption}, we compare our
theoretical prediction with the experimental points of Long Ju and
coworkers \citep{ju2017tunable}. To facilitate the comparison between
theory and experiment, we chose a bias of $V=52$ meV since it produced
the best description of the experimental results; the remaining parameters
retain the same values we considered so far (with $r_0$ taken from \citep{Li2019}). Comparing the theoretical
prediction with the experimental data an excellent agreement is seen,
in both the position and the magnitude of the two resonances. Besides
this, we also see that the theoretical plateau at higher energies
matches well with the experimental one. Although our plateau appears
to be further away from the 2p resonance than in the experimental
data, it seems possible to assign the structures that are experimentally
captured on the high energy side of the largest peak to other excitonic state, which, due to their proximity, merge together, creating a single broader resonance.

\begin{figure}[h]
\centering{}\includegraphics{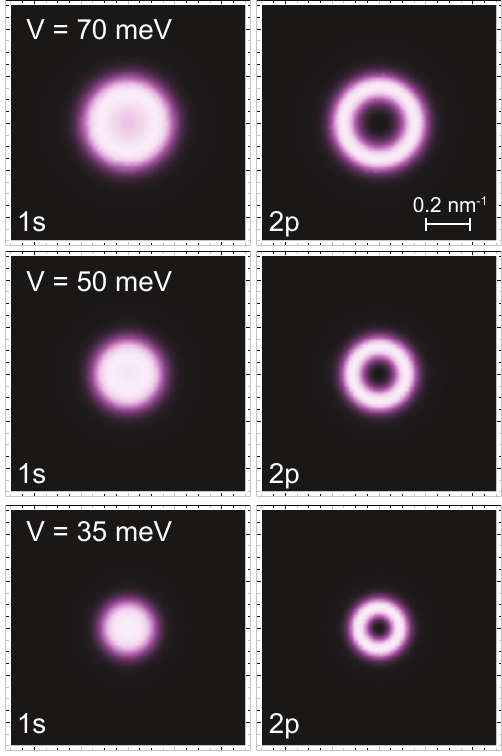}\caption{\label{fig:WF_Panel}Density plot of the absolute value of the wavefunctions
of the 1s and 2p states for three bias values $V=35$ meV, $V=50$
meV and $V=70$ meV. The same scaled is used in all panels.}
\end{figure}

At last, in Fig. \ref{fig:WF_Panel}, we depict a density plot of
the $1s$ and $2p$ exciton wave functions for different bias. 
As expected, for every case the $1s$ wave function is finite at $k=0$
while it vanishes for the $2p$ states. Moreover, as the bias increases
the wave functions become more delocalized in momentum space. This
increased delocalization in momentum space, translates to a higher
localization in real space, in agreement with the aforementioned idea
that the excitons become more tightly bound as the bias rises. This
relation between real and momentum space also allows us to estimate
the size of an exciton. Taking the case of $V=70$ meV as an example,
we see that its $2p$ wave function presents a spread in momentum
space of about $\Delta k=0.2$ nm$^{-1}$. Using $\Delta r=2\pi/\Delta k$,
we estimate that the same exciton should be extended over approximately
40 nm in real space.

\section{Conclusion\label{sec:Conclusion}}

In this paper we studied the excitonic response of biased bilayer
graphene. To achieve this, we started by describing the single particle
electronic properties of the system using a tight binding Hamiltonian.
Afterwards, the single particle solutions were used as the input to
the Bethe-Salpeter equation, whose solution defines the excitonic
properties, i.e. their energies and wave functions. To solve the BSE
we transformed the 2D integral equation into a 1D problem by carefully
choosing the phase of the Bloch factors. This reduction in the dimension
of the problem allows us to more efficiently solve the BSE, which
now requires a single numerical quadrature.

To study the optical properties of the biased bilayer, we computed
its conductivity, which allowed us to extract the optical selection
rules and gain intuition on the strength of the photon-exciton coupling.
We found that, in agreement with the literature, when trigonal warping
is ignored, only $p-$states (with $m=\pm1$) are bright; even though
$f-$states are optically allowed they present a tiny oscillator strength.
When trigonal warping is considered, and thus the symmetry of the
calculation is reduced from $C_{\infty}$ to $C_{3}$, we find that
$s-$states (with $m=0$) become optically allowed, and present a
noticeable oscillator strength (at least for the 1s state).

Making use of the optical conductivity, we computed the absorption
spectrum for the case where a biased bilayer graphene is encapsulated
in hBN, which corresponds to the system where an experimental study
was carried out by Long Ju and coworkers in \citep{ju2017tunable}.
Comparing our theoretical prediction with the experimental data, an
excellent agreement was observed. Our model accurately described the
position and magnitude of the 1s and 2p resonances, while also capturing
the transition from excitonic resonances to the continuum of states,
which originates the plateau that appears in the experimentally measured
absorption. The dependence of the resonance's location with the bias
also agreed with the experimental data.

\emph{Note added: After the completion of this work we became aware
of a similar unpublished work \citep{sauer2021exciton} on the same
system we treated in this paper.}

\begin{widetext}

\appendix

\section{On the solution of the Bethe-Salpeter equation}

In this appendix we shall give a more in depth description on how
to numerically solve the Bethe-Salpeter equation (BSE). This approach
is an extension of the method presented in Ref. Ref. \citep{Chao1991}
to treat excitons in 2D semiconductors using the Coulomb potential.
We take Eq. (\ref{eq:BSE}) of the main text as our starting point.
Using the \emph{ansatz }$\psi_{cv}(\mathbf{k})=f_{cv}(k)e^{im\theta}$,
and taking the thermodynamic limit, we find:
\begin{equation}
\left(E_{k}^{c}-E_{k}^{v}\right)f_{cv}(k)-\frac{1}{4\pi^{2}}\sum_{c'v'}\int qdqd\theta_{q}V(\mathbf{k}-\mathbf{q})\langle u_{\mathbf{k}}^{c}|u_{\mathbf{q}}^{c'}\rangle\langle u_{\mathbf{q}}^{v'}|u_{\mathbf{k}}^{v}\rangle f_{c'v'}(q)e^{im\left(\theta_{q}-\theta_{k}\right)}=Ef_{cv}(k).
\end{equation}
This equation corresponds to a two dimensional integral equation whose
numerical solution is computationally demanding. To reduce the numerical
weight of the calculation we wish to reduce the problem to that of
a one dimensional integral equation. This can be achieved if the product
of Bloch factors has the form
\begin{equation}
\langle u_{\mathbf{k}}^{c}|u_{\mathbf{q}}^{c'}\rangle\langle u_{\mathbf{q}}^{v'}|u_{\mathbf{k}}^{v}\rangle=\sum_{\lambda}\mathcal{A}_{\lambda}^{cc'vv'}(k,q)e^{i\lambda(\theta_{q}-\theta_{k})},
\end{equation}
since this allows us to perform a variable change from to $\vartheta=\theta_{q}-\theta_{k}$,
thus eliminating one of the momentum dependent integrals; here $\lambda$
is some integer, and $\mathcal{A}_{\lambda}^{cc'vv'}(k,q)$ are coefficients
determined by the explicit computation of the spinor product. In principle
this is always possible for the materials we are considering, given
that the phases of the Bloch factors are appropriately chosen. Inserting
this into the previous equation, and noting that $V(\mathbf{k}-\mathbf{q})\equiv V(k,q,\theta_{q}-\theta_{k})$,
one finds
\begin{equation}
\left(E_{k}^{c}-E_{k}^{v}\right)f_{cv}(k)-\frac{1}{4\pi^{2}}\sum_{c'v'}\sum_{\lambda}\int qdqd\vartheta V(k,q,\vartheta)\mathcal{A}_{\lambda}^{cc'vv'}(k,q)f_{c'v'}(q)e^{i\left(m+\lambda\right)\vartheta}=Ef_{cv}(k),
\end{equation}
where we explicitly introduced the variable change $d\theta_{q}\rightarrow d\vartheta$.
Now, recalling the definition of Rytova-Keldysh potential $V(k,q,\vartheta)$,
we introduce a new function, $\mathcal{I}_{\nu}(k,q)$ corresponding
to the integral over $d\vartheta$:
\begin{equation}
\mathcal{I}_{\nu}(k,q)=\int_{0}^{2\pi}\frac{\cos\left(\nu\vartheta\right)}{\kappa(k,q,\vartheta)\left[1+r_{0}\kappa(k,q,\vartheta)\right]}d\vartheta
\end{equation}
with $\kappa(k,q,\vartheta)=\sqrt{k^{2}+q^{2}-2kq\cos\vartheta}$.
Notice how only $\cos(\nu\theta)$ enters the integral, since the
analogous term in $\sin(\nu\vartheta)$ vanishes due to parity. From
inspection, it should be clear that when $q=k$ the function $\mathcal{I}_{\nu}(k,q)$
is numerically ill-behaved, and as such must be treated carefully.
For convenience we shall express $\mathcal{I}_{\nu}(k,q)$ in terms
of partial fractions as
\begin{align}
\mathcal{I}_{\nu}(k,q) & =\int_{0}^{2\pi}\frac{\cos\left(\nu\vartheta\right)}{\kappa(k,q,\vartheta)}d\vartheta-r_{0}\int_{0}^{2\pi}\frac{\cos\left(\nu\vartheta\right)}{\left[1+r_{0}\kappa(k,q,\vartheta)\right]}d\vartheta\\
 & \equiv\mathcal{J}_{\nu}(k,q)-\mathcal{K}_{\nu}(k,q),
\end{align}
where from these two terms only the first one, $\mathcal{J}_{\nu}(k,q)$,
is problematic when $k=q$. Before we explain how to avoid this numerical
problem, let us first write the BSE in a more useful way. First, we
write
\begin{equation}
\left(E_{k}^{c}-E_{k}^{v}\right)f_{cv}(k)-\sum_{c'v'}\sum_{\lambda}\int_{0}^{\infty}\left\{ \mathcal{J}_{m+\lambda}(k,q)\mathcal{A}_{\lambda}^{cc'vv'}(k,q)f_{c'v'}(q)-\mathcal{K}_{m+\lambda}(k,q)\mathcal{A}_{\lambda}^{cc'vv'}(k,q)f_{c'v'}(q)\right\} qdq=Ef_{cv}(k).
\end{equation}
Then, we define $\mathcal{B}_{m}^{cc'vv'}(k,q)=\sum_{\lambda}\mathcal{J}_{m+\lambda}(k,q)\mathcal{A}_{\lambda}^{cc'vv'}(k,q)$
and $\mathcal{C}_{m}^{cc'vv'}(k,q)=\sum_{\lambda}\mathcal{K}_{m+\lambda}(k,q)\mathcal{A}_{\lambda}^{cc'vv'}(k,q)$.
Using these, one finds
\begin{equation}
\left(E_{k}^{c}-E_{k}^{v}\right)f_{cv}(k)-\sum_{c'v'}\int_{0}^{\infty}\mathcal{B}_{m}^{cc'vv'}(k,q)f_{c'v'}(q)qdq+\sum_{c'v'}\int_{0}^{\infty}\mathcal{C}_{m}^{cc'vv'}(k,q)f_{c'v'}(q)qdq=Ef_{cv}(k).
\end{equation}

Let us now focus on the numerical problem associated with $\mathcal{B}_{m}^{cc'vv'}(k,q)$.
To treat the divergence that appears when $k=q$, we introduce an
auxiliary function $g_{m}(k,q)$ and introduce the modification
\begin{equation}
\int_{0}^{\infty}\mathcal{B}_{m}^{cc'vv'}(k,q)f_{c'v'}(q)qdq\rightarrow\int_{0}^{\infty}\left[\mathcal{B}_{m}^{cc'vv'}(k,q)f_{c'v'}(q)-g_{m}(k,q)f_{c'v'}(k)\right]qdq+f_{c'v'}(k)\int_{0}^{\infty}g_{m}(k,q)qdq,
\end{equation}
with $g_{m}$ defined such that $\lim_{q\rightarrow k}\left[\mathcal{B}_{m}^{cc'vv'}(k,q)-g_{m}(k,q)\right]=0$.
Following Ref. \citep{Chao1991}, we define $g_{m}$ as
\begin{equation}
g_{m}=\mathcal{B}_{m}^{cc'vv'}(k,q)\frac{2k^{2}}{k^{2}+q^{2}}
\end{equation}

With the analytical part of the calculation taken care of, we shall
now discuss how to numerically solve the equation we have arrived
to. First, we introduce a variable change which transforms the improper
integral over $[0,\infty)$, into one with finite integration limits,
such as $[0,1]$. One possibility is to define $q=\tan[\pi x/2]$.
Afterwards, we discretize the variables $k$ and $x$ (and consequently
$q$):
\begin{align}
 & \left(E_{k_{i}}^{c}-E_{k_{i}}^{v}\right)f_{cv}(k_{i})+\sum_{c'v'}\sum_{j=1}^{N}\mathcal{C}_{m}^{cc'vv'}(k_{i},q_{j})f_{c'v'}(q_{j})q_{j}\frac{dq}{dx_{j}}\nonumber \\
 & -\sum_{c'v'}\sum_{j\neq i}\mathcal{B}_{m}^{cc'vv'}(k_{i},q_{j})f_{c'v'}(q_{j})q_{j}\frac{dq}{dx_{j}}w_{j}-f_{c'v'}(k_{i})\left\{ \int_{0}^{\infty}g_{m}(k_{i},p)pdp-\sum_{j\neq i}g_{m}(k_{i},q_{j})q_{j}\frac{dq}{dx_{j}}\right\} =Ef_{cv}(k_{i})
\end{align}
where $N$ is the number of points and $w_{j}$ is the weight function
of the chosen numerical quadrature; also, $q_{j}\equiv q(x_{j})$
and $dq/dx_{j}\equiv\left[dq/dx\right]_{x=x_{j}}$. Furthermore, we
note that $\int_{0}^{\infty}g_{m}(k_{i},p)pdp$ is numerically well
behaved as opposed to the original integral, $\int_{0}^{\infty}\mathcal{B}_{m}^{cc'vv'}(k_{i},p)pdp$.
Depending on the problem, and on how localized the excitonic wave
functions are in momentum space, it may be useful to split the original
improper integral into regions, in order to allow for a thinner mesh
in the relevant portion of $k$-space, and a coarser one in the region
where the wave functions are already close to zero.

Regarding the choice of quadrature, we employ a Gauss-Legendre quadrature,
which is defined as \citep{kythe2011computational}
\begin{equation}
\int_{a}^{b}f(y)dy\approx\sum_{i=1}^{N}w_{i}f(y_{i})
\end{equation}
where 
\begin{equation}
y_{i}=\frac{a+b+(b-a)\xi_{i}}{2},
\end{equation}
with $\xi_{i}$ the $i-$th zero of the Legendre polynomial $P_{N}(y)$,
and
\begin{equation}
w_{i}=\frac{b-a}{(1-\xi_{i})^{2}\left[P'_{N}(\xi_{i})\right]^{2}}
\end{equation}
with $P'_{N}(\xi_{i})\equiv\left[dP_{N}(y)/dy\right]_{y=\xi_{i}}$.

At last, the only thing left to do is to realize that this equation
can be expressed as an eigenvalue problem of a $4N\times4N$ matrix.
This matrix can be thought of as a $4\times4$ matrix of matrices,
each one with dimensions $N\times N$. The 16 blocks come from the
different combinations of the indexes $c$, $c'$, $v$ and $v'$,
with each block corresponding to a $N\times N$ matrix stemming from
the numerical discretization of the integral. Solving the eigenvalue
problem one finds the exciton energies and wave functions.\end{widetext}

\bibliographystyle{apsrev4-2}

\begin{thebibliography}{33}%
\makeatletter
\providecommand \@ifxundefined [1]{%
 \@ifx{#1\undefined}
}%
\providecommand \@ifnum [1]{%
 \ifnum #1\expandafter \@firstoftwo
 \else \expandafter \@secondoftwo
 \fi
}%
\providecommand \@ifx [1]{%
 \ifx #1\expandafter \@firstoftwo
 \else \expandafter \@secondoftwo
 \fi
}%
\providecommand \natexlab [1]{#1}%
\providecommand \enquote  [1]{``#1''}%
\providecommand \bibnamefont  [1]{#1}%
\providecommand \bibfnamefont [1]{#1}%
\providecommand \citenamefont [1]{#1}%
\providecommand \href@noop [0]{\@secondoftwo}%
\providecommand \href [0]{\begingroup \@sanitize@url \@href}%
\providecommand \@href[1]{\@@startlink{#1}\@@href}%
\providecommand \@@href[1]{\endgroup#1\@@endlink}%
\providecommand \@sanitize@url [0]{\catcode `\\12\catcode `\$12\catcode
  `\&12\catcode `\#12\catcode `\^12\catcode `\_12\catcode `\%12\relax}%
\providecommand \@@startlink[1]{}%
\providecommand \@@endlink[0]{}%
\providecommand \url  [0]{\begingroup\@sanitize@url \@url }%
\providecommand \@url [1]{\endgroup\@href {#1}{\urlprefix }}%
\providecommand \urlprefix  [0]{URL }%
\providecommand \Eprint [0]{\href }%
\providecommand \doibase [0]{https://doi.org/}%
\providecommand \selectlanguage [0]{\@gobble}%
\providecommand \bibinfo  [0]{\@secondoftwo}%
\providecommand \bibfield  [0]{\@secondoftwo}%
\providecommand \translation [1]{[#1]}%
\providecommand \BibitemOpen [0]{}%
\providecommand \bibitemStop [0]{}%
\providecommand \bibitemNoStop [0]{.\EOS\space}%
\providecommand \EOS [0]{\spacefactor3000\relax}%
\providecommand \BibitemShut  [1]{\csname bibitem#1\endcsname}%
\let\auto@bib@innerbib\@empty
\bibitem [{\citenamefont {Novoselov}\ \emph {et~al.}(2004)\citenamefont
  {Novoselov}, \citenamefont {Geim}, \citenamefont {Morozov}, \citenamefont
  {Jiang}, \citenamefont {Zhang}, \citenamefont {Dubonos}, \citenamefont
  {Grigorieva},\ and\ \citenamefont {Firsov}}]{novoselov2004electric}%
  \BibitemOpen
  \bibfield  {author} {\bibinfo {author} {\bibfnamefont {K.~S.}\ \bibnamefont
  {Novoselov}}, \bibinfo {author} {\bibfnamefont {A.~K.}\ \bibnamefont {Geim}},
  \bibinfo {author} {\bibfnamefont {S.~V.}\ \bibnamefont {Morozov}}, \bibinfo
  {author} {\bibfnamefont {D.-e.}\ \bibnamefont {Jiang}}, \bibinfo {author}
  {\bibfnamefont {Y.}~\bibnamefont {Zhang}}, \bibinfo {author} {\bibfnamefont
  {S.~V.}\ \bibnamefont {Dubonos}}, \bibinfo {author} {\bibfnamefont {I.~V.}\
  \bibnamefont {Grigorieva}},\ and\ \bibinfo {author} {\bibfnamefont {A.~A.}\
  \bibnamefont {Firsov}},\ }\href@noop {} {\bibfield  {journal} {\bibinfo
  {journal} {science}\ }\textbf {\bibinfo {volume} {306}},\ \bibinfo {pages}
  {666} (\bibinfo {year} {2004})}\BibitemShut {NoStop}%
\bibitem [{\citenamefont {Wang}\ \emph {et~al.}(2018)\citenamefont {Wang},
  \citenamefont {Chernikov}, \citenamefont {Glazov}, \citenamefont {Heinz},
  \citenamefont {Marie}, \citenamefont {Amand},\ and\ \citenamefont
  {Urbaszek}}]{Wang2018Colloquium}%
  \BibitemOpen
  \bibfield  {author} {\bibinfo {author} {\bibfnamefont {G.}~\bibnamefont
  {Wang}}, \bibinfo {author} {\bibfnamefont {A.}~\bibnamefont {Chernikov}},
  \bibinfo {author} {\bibfnamefont {M.~M.}\ \bibnamefont {Glazov}}, \bibinfo
  {author} {\bibfnamefont {T.~F.}\ \bibnamefont {Heinz}}, \bibinfo {author}
  {\bibfnamefont {X.}~\bibnamefont {Marie}}, \bibinfo {author} {\bibfnamefont
  {T.}~\bibnamefont {Amand}},\ and\ \bibinfo {author} {\bibfnamefont
  {B.}~\bibnamefont {Urbaszek}},\ }\href
  {https://doi.org/10.1103/RevModPhys.90.021001} {\bibfield  {journal}
  {\bibinfo  {journal} {Rev. Mod. Phys.}\ }\textbf {\bibinfo {volume} {90}},\
  \bibinfo {pages} {021001} (\bibinfo {year} {2018})}\BibitemShut {NoStop}%
\bibitem [{\citenamefont {Carvalho}\ \emph {et~al.}(2016)\citenamefont
  {Carvalho}, \citenamefont {Wang}, \citenamefont {Zhu}, \citenamefont {Rodin},
  \citenamefont {Su},\ and\ \citenamefont {Neto}}]{carvalho2016phosphorene}%
  \BibitemOpen
  \bibfield  {author} {\bibinfo {author} {\bibfnamefont {A.}~\bibnamefont
  {Carvalho}}, \bibinfo {author} {\bibfnamefont {M.}~\bibnamefont {Wang}},
  \bibinfo {author} {\bibfnamefont {X.}~\bibnamefont {Zhu}}, \bibinfo {author}
  {\bibfnamefont {A.~S.}\ \bibnamefont {Rodin}}, \bibinfo {author}
  {\bibfnamefont {H.}~\bibnamefont {Su}},\ and\ \bibinfo {author}
  {\bibfnamefont {A.~H.~C.}\ \bibnamefont {Neto}},\ }\href@noop {} {\bibfield
  {journal} {\bibinfo  {journal} {Nature Reviews Materials}\ }\textbf {\bibinfo
  {volume} {1}},\ \bibinfo {pages} {1} (\bibinfo {year} {2016})}\BibitemShut
  {NoStop}%
\bibitem [{\citenamefont {Caldwell}\ \emph {et~al.}(2019)\citenamefont
  {Caldwell}, \citenamefont {Aharonovich}, \citenamefont {Cassabois},
  \citenamefont {Edgar}, \citenamefont {Gil},\ and\ \citenamefont
  {Basov}}]{caldwell2019photonics}%
  \BibitemOpen
  \bibfield  {author} {\bibinfo {author} {\bibfnamefont {J.~D.}\ \bibnamefont
  {Caldwell}}, \bibinfo {author} {\bibfnamefont {I.}~\bibnamefont
  {Aharonovich}}, \bibinfo {author} {\bibfnamefont {G.}~\bibnamefont
  {Cassabois}}, \bibinfo {author} {\bibfnamefont {J.~H.}\ \bibnamefont
  {Edgar}}, \bibinfo {author} {\bibfnamefont {B.}~\bibnamefont {Gil}},\ and\
  \bibinfo {author} {\bibfnamefont {D.}~\bibnamefont {Basov}},\ }\href@noop {}
  {\bibfield  {journal} {\bibinfo  {journal} {Nature Reviews Materials}\
  }\textbf {\bibinfo {volume} {4}},\ \bibinfo {pages} {552} (\bibinfo {year}
  {2019})}\BibitemShut {NoStop}%
\bibitem [{\citenamefont {Cudazzo}\ \emph {et~al.}(2011)\citenamefont
  {Cudazzo}, \citenamefont {Tokatly},\ and\ \citenamefont
  {Rubio}}]{Cudazzo2011}%
  \BibitemOpen
  \bibfield  {author} {\bibinfo {author} {\bibfnamefont {P.}~\bibnamefont
  {Cudazzo}}, \bibinfo {author} {\bibfnamefont {I.~V.}\ \bibnamefont
  {Tokatly}},\ and\ \bibinfo {author} {\bibfnamefont {A.}~\bibnamefont
  {Rubio}},\ }\href {https://doi.org/10.1103/PhysRevB.84.085406} {\bibfield
  {journal} {\bibinfo  {journal} {Phys. Rev. B}\ }\textbf {\bibinfo {volume}
  {84}},\ \bibinfo {pages} {085406} (\bibinfo {year} {2011})}\BibitemShut
  {NoStop}%
\bibitem [{\citenamefont {Watanabe}\ \emph {et~al.}(2004)\citenamefont
  {Watanabe}, \citenamefont {Taniguchi},\ and\ \citenamefont
  {Kanda}}]{watanabe2004direct}%
  \BibitemOpen
  \bibfield  {author} {\bibinfo {author} {\bibfnamefont {K.}~\bibnamefont
  {Watanabe}}, \bibinfo {author} {\bibfnamefont {T.}~\bibnamefont
  {Taniguchi}},\ and\ \bibinfo {author} {\bibfnamefont {H.}~\bibnamefont
  {Kanda}},\ }\href@noop {} {\bibfield  {journal} {\bibinfo  {journal} {Nature
  materials}\ }\textbf {\bibinfo {volume} {3}},\ \bibinfo {pages} {404}
  (\bibinfo {year} {2004})}\BibitemShut {NoStop}%
\bibitem [{\citenamefont {Kubota}\ \emph {et~al.}(2007)\citenamefont {Kubota},
  \citenamefont {Watanabe}, \citenamefont {Tsuda},\ and\ \citenamefont
  {Taniguchi}}]{kubota2007deep}%
  \BibitemOpen
  \bibfield  {author} {\bibinfo {author} {\bibfnamefont {Y.}~\bibnamefont
  {Kubota}}, \bibinfo {author} {\bibfnamefont {K.}~\bibnamefont {Watanabe}},
  \bibinfo {author} {\bibfnamefont {O.}~\bibnamefont {Tsuda}},\ and\ \bibinfo
  {author} {\bibfnamefont {T.}~\bibnamefont {Taniguchi}},\ }\href@noop {}
  {\bibfield  {journal} {\bibinfo  {journal} {Science}\ }\textbf {\bibinfo
  {volume} {317}},\ \bibinfo {pages} {932} (\bibinfo {year}
  {2007})}\BibitemShut {NoStop}%
\bibitem [{\citenamefont {Schaibley}\ \emph {et~al.}(2016)\citenamefont
  {Schaibley}, \citenamefont {Yu}, \citenamefont {Clark}, \citenamefont
  {Rivera}, \citenamefont {Ross}, \citenamefont {Seyler}, \citenamefont {Yao},\
  and\ \citenamefont {Xu}}]{schaibley2016valleytronics}%
  \BibitemOpen
  \bibfield  {author} {\bibinfo {author} {\bibfnamefont {J.~R.}\ \bibnamefont
  {Schaibley}}, \bibinfo {author} {\bibfnamefont {H.}~\bibnamefont {Yu}},
  \bibinfo {author} {\bibfnamefont {G.}~\bibnamefont {Clark}}, \bibinfo
  {author} {\bibfnamefont {P.}~\bibnamefont {Rivera}}, \bibinfo {author}
  {\bibfnamefont {J.~S.}\ \bibnamefont {Ross}}, \bibinfo {author}
  {\bibfnamefont {K.~L.}\ \bibnamefont {Seyler}}, \bibinfo {author}
  {\bibfnamefont {W.}~\bibnamefont {Yao}},\ and\ \bibinfo {author}
  {\bibfnamefont {X.}~\bibnamefont {Xu}},\ }\href@noop {} {\bibfield  {journal}
  {\bibinfo  {journal} {Nature Reviews Materials}\ }\textbf {\bibinfo {volume}
  {1}},\ \bibinfo {pages} {1} (\bibinfo {year} {2016})}\BibitemShut {NoStop}%
\bibitem [{\citenamefont {Yu}\ \emph {et~al.}(2017)\citenamefont {Yu},
  \citenamefont {Liu}, \citenamefont {Tang}, \citenamefont {Xu},\ and\
  \citenamefont {Yao}}]{yu2017moire}%
  \BibitemOpen
  \bibfield  {author} {\bibinfo {author} {\bibfnamefont {H.}~\bibnamefont
  {Yu}}, \bibinfo {author} {\bibfnamefont {G.-B.}\ \bibnamefont {Liu}},
  \bibinfo {author} {\bibfnamefont {J.}~\bibnamefont {Tang}}, \bibinfo {author}
  {\bibfnamefont {X.}~\bibnamefont {Xu}},\ and\ \bibinfo {author}
  {\bibfnamefont {W.}~\bibnamefont {Yao}},\ }\href@noop {} {\bibfield
  {journal} {\bibinfo  {journal} {Science advances}\ }\textbf {\bibinfo
  {volume} {3}},\ \bibinfo {pages} {e1701696} (\bibinfo {year}
  {2017})}\BibitemShut {NoStop}%
\bibitem [{\citenamefont {Branny}\ \emph {et~al.}(2017)\citenamefont {Branny},
  \citenamefont {Kumar}, \citenamefont {Proux},\ and\ \citenamefont
  {Gerardot}}]{branny2017deterministic}%
  \BibitemOpen
  \bibfield  {author} {\bibinfo {author} {\bibfnamefont {A.}~\bibnamefont
  {Branny}}, \bibinfo {author} {\bibfnamefont {S.}~\bibnamefont {Kumar}},
  \bibinfo {author} {\bibfnamefont {R.}~\bibnamefont {Proux}},\ and\ \bibinfo
  {author} {\bibfnamefont {B.~D.}\ \bibnamefont {Gerardot}},\ }\href@noop {}
  {\bibfield  {journal} {\bibinfo  {journal} {Nature communications}\ }\textbf
  {\bibinfo {volume} {8}},\ \bibinfo {pages} {1} (\bibinfo {year}
  {2017})}\BibitemShut {NoStop}%
\bibitem [{\citenamefont {Epstein}\ \emph {et~al.}(2020)\citenamefont
  {Epstein}, \citenamefont {Terr{\'e}s}, \citenamefont {Chaves}, \citenamefont
  {Pusapati}, \citenamefont {Rhodes}, \citenamefont {Frank}, \citenamefont
  {Zimmermann}, \citenamefont {Qin}, \citenamefont {Watanabe}, \citenamefont
  {Taniguchi}, \citenamefont {Giessen}, \citenamefont {Tongay}, \citenamefont
  {Hone}, \citenamefont {Peres},\ and\ \citenamefont
  {Koppens}}]{epstein2020near}%
  \BibitemOpen
  \bibfield  {author} {\bibinfo {author} {\bibfnamefont {I.}~\bibnamefont
  {Epstein}}, \bibinfo {author} {\bibfnamefont {B.}~\bibnamefont {Terr{\'e}s}},
  \bibinfo {author} {\bibfnamefont {A.~J.}\ \bibnamefont {Chaves}}, \bibinfo
  {author} {\bibfnamefont {V.-V.}\ \bibnamefont {Pusapati}}, \bibinfo {author}
  {\bibfnamefont {D.~A.}\ \bibnamefont {Rhodes}}, \bibinfo {author}
  {\bibfnamefont {B.}~\bibnamefont {Frank}}, \bibinfo {author} {\bibfnamefont
  {V.}~\bibnamefont {Zimmermann}}, \bibinfo {author} {\bibfnamefont
  {Y.}~\bibnamefont {Qin}}, \bibinfo {author} {\bibfnamefont {K.}~\bibnamefont
  {Watanabe}}, \bibinfo {author} {\bibfnamefont {T.}~\bibnamefont {Taniguchi}},
  \bibinfo {author} {\bibfnamefont {H.}~\bibnamefont {Giessen}}, \bibinfo
  {author} {\bibfnamefont {S.}~\bibnamefont {Tongay}}, \bibinfo {author}
  {\bibfnamefont {J.~C.}\ \bibnamefont {Hone}}, \bibinfo {author}
  {\bibfnamefont {N.~M.~R.}\ \bibnamefont {Peres}},\ and\ \bibinfo {author}
  {\bibfnamefont {F.~H.~L.}\ \bibnamefont {Koppens}},\ }\href@noop {}
  {\bibfield  {journal} {\bibinfo  {journal} {Nano letters}\ }\textbf {\bibinfo
  {volume} {20}},\ \bibinfo {pages} {3545} (\bibinfo {year}
  {2020})}\BibitemShut {NoStop}%
\bibitem [{\citenamefont {Castro}\ \emph {et~al.}(2007)\citenamefont {Castro},
  \citenamefont {Novoselov}, \citenamefont {Morozov}, \citenamefont {Peres},
  \citenamefont {Dos~Santos}, \citenamefont {Nilsson}, \citenamefont {Guinea},
  \citenamefont {Geim},\ and\ \citenamefont {Neto}}]{castro2007biased}%
  \BibitemOpen
  \bibfield  {author} {\bibinfo {author} {\bibfnamefont {E.~V.}\ \bibnamefont
  {Castro}}, \bibinfo {author} {\bibfnamefont {K.}~\bibnamefont {Novoselov}},
  \bibinfo {author} {\bibfnamefont {S.}~\bibnamefont {Morozov}}, \bibinfo
  {author} {\bibfnamefont {N.}~\bibnamefont {Peres}}, \bibinfo {author}
  {\bibfnamefont {J.~L.}\ \bibnamefont {Dos~Santos}}, \bibinfo {author}
  {\bibfnamefont {J.}~\bibnamefont {Nilsson}}, \bibinfo {author} {\bibfnamefont
  {F.}~\bibnamefont {Guinea}}, \bibinfo {author} {\bibfnamefont
  {A.}~\bibnamefont {Geim}},\ and\ \bibinfo {author} {\bibfnamefont {A.~C.}\
  \bibnamefont {Neto}},\ }\href@noop {} {\bibfield  {journal} {\bibinfo
  {journal} {Physical review letters}\ }\textbf {\bibinfo {volume} {99}},\
  \bibinfo {pages} {216802} (\bibinfo {year} {2007})}\BibitemShut {NoStop}%
\bibitem [{\citenamefont {Zhang}\ \emph {et~al.}(2009)\citenamefont {Zhang},
  \citenamefont {Tang}, \citenamefont {Girit}, \citenamefont {Hao},
  \citenamefont {Martin}, \citenamefont {Zettl}, \citenamefont {Crommie},
  \citenamefont {Shen},\ and\ \citenamefont {Wang}}]{zhang2009direct}%
  \BibitemOpen
  \bibfield  {author} {\bibinfo {author} {\bibfnamefont {Y.}~\bibnamefont
  {Zhang}}, \bibinfo {author} {\bibfnamefont {T.-T.}\ \bibnamefont {Tang}},
  \bibinfo {author} {\bibfnamefont {C.}~\bibnamefont {Girit}}, \bibinfo
  {author} {\bibfnamefont {Z.}~\bibnamefont {Hao}}, \bibinfo {author}
  {\bibfnamefont {M.~C.}\ \bibnamefont {Martin}}, \bibinfo {author}
  {\bibfnamefont {A.}~\bibnamefont {Zettl}}, \bibinfo {author} {\bibfnamefont
  {M.~F.}\ \bibnamefont {Crommie}}, \bibinfo {author} {\bibfnamefont {Y.~R.}\
  \bibnamefont {Shen}},\ and\ \bibinfo {author} {\bibfnamefont
  {F.}~\bibnamefont {Wang}},\ }\href@noop {} {\bibfield  {journal} {\bibinfo
  {journal} {Nature}\ }\textbf {\bibinfo {volume} {459}},\ \bibinfo {pages}
  {820} (\bibinfo {year} {2009})}\BibitemShut {NoStop}%
\bibitem [{\citenamefont {Oostinga}\ \emph {et~al.}(2008)\citenamefont
  {Oostinga}, \citenamefont {Heersche}, \citenamefont {Liu}, \citenamefont
  {Morpurgo},\ and\ \citenamefont {Vandersypen}}]{oostinga2008gate}%
  \BibitemOpen
  \bibfield  {author} {\bibinfo {author} {\bibfnamefont {J.~B.}\ \bibnamefont
  {Oostinga}}, \bibinfo {author} {\bibfnamefont {H.~B.}\ \bibnamefont
  {Heersche}}, \bibinfo {author} {\bibfnamefont {X.}~\bibnamefont {Liu}},
  \bibinfo {author} {\bibfnamefont {A.~F.}\ \bibnamefont {Morpurgo}},\ and\
  \bibinfo {author} {\bibfnamefont {L.~M.}\ \bibnamefont {Vandersypen}},\
  }\href@noop {} {\bibfield  {journal} {\bibinfo  {journal} {Nature materials}\
  }\textbf {\bibinfo {volume} {7}},\ \bibinfo {pages} {151} (\bibinfo {year}
  {2008})}\BibitemShut {NoStop}%
\bibitem [{\citenamefont {Park}\ and\ \citenamefont
  {Louie}(2010)}]{park2010tunable}%
  \BibitemOpen
  \bibfield  {author} {\bibinfo {author} {\bibfnamefont {C.-H.}\ \bibnamefont
  {Park}}\ and\ \bibinfo {author} {\bibfnamefont {S.~G.}\ \bibnamefont
  {Louie}},\ }\href@noop {} {\bibfield  {journal} {\bibinfo  {journal} {Nano
  letters}\ }\textbf {\bibinfo {volume} {10}},\ \bibinfo {pages} {426}
  (\bibinfo {year} {2010})}\BibitemShut {NoStop}%
\bibitem [{\citenamefont {Ju}\ \emph {et~al.}(2017)\citenamefont {Ju},
  \citenamefont {Wang}, \citenamefont {Cao}, \citenamefont {Taniguchi},
  \citenamefont {Watanabe}, \citenamefont {Louie}, \citenamefont {Rana},
  \citenamefont {Park}, \citenamefont {Hone}, \citenamefont {Wang} \emph
  {et~al.}}]{ju2017tunable}%
  \BibitemOpen
  \bibfield  {author} {\bibinfo {author} {\bibfnamefont {L.}~\bibnamefont
  {Ju}}, \bibinfo {author} {\bibfnamefont {L.}~\bibnamefont {Wang}}, \bibinfo
  {author} {\bibfnamefont {T.}~\bibnamefont {Cao}}, \bibinfo {author}
  {\bibfnamefont {T.}~\bibnamefont {Taniguchi}}, \bibinfo {author}
  {\bibfnamefont {K.}~\bibnamefont {Watanabe}}, \bibinfo {author}
  {\bibfnamefont {S.~G.}\ \bibnamefont {Louie}}, \bibinfo {author}
  {\bibfnamefont {F.}~\bibnamefont {Rana}}, \bibinfo {author} {\bibfnamefont
  {J.}~\bibnamefont {Park}}, \bibinfo {author} {\bibfnamefont {J.}~\bibnamefont
  {Hone}}, \bibinfo {author} {\bibfnamefont {F.}~\bibnamefont {Wang}}, \emph
  {et~al.},\ }\href@noop {} {\bibfield  {journal} {\bibinfo  {journal}
  {Science}\ }\textbf {\bibinfo {volume} {358}},\ \bibinfo {pages} {907}
  (\bibinfo {year} {2017})}\BibitemShut {NoStop}%
\bibitem [{\citenamefont {Fuchs}\ \emph {et~al.}(2008)\citenamefont {Fuchs},
  \citenamefont {R\"odl}, \citenamefont {Schleife},\ and\ \citenamefont
  {Bechstedt}}]{Fuchs2013}%
  \BibitemOpen
  \bibfield  {author} {\bibinfo {author} {\bibfnamefont {F.}~\bibnamefont
  {Fuchs}}, \bibinfo {author} {\bibfnamefont {C.}~\bibnamefont {R\"odl}},
  \bibinfo {author} {\bibfnamefont {A.}~\bibnamefont {Schleife}},\ and\
  \bibinfo {author} {\bibfnamefont {F.}~\bibnamefont {Bechstedt}},\ }\href
  {https://doi.org/10.1103/PhysRevB.78.085103} {\bibfield  {journal} {\bibinfo
  {journal} {Phys. Rev. B}\ }\textbf {\bibinfo {volume} {78}},\ \bibinfo
  {pages} {085103} (\bibinfo {year} {2008})}\BibitemShut {NoStop}%
\bibitem [{\citenamefont {Komsa}\ and\ \citenamefont
  {Krasheninnikov}(2013)}]{Komsa2013}%
  \BibitemOpen
  \bibfield  {author} {\bibinfo {author} {\bibfnamefont {H.-P.}\ \bibnamefont
  {Komsa}}\ and\ \bibinfo {author} {\bibfnamefont {A.~V.}\ \bibnamefont
  {Krasheninnikov}},\ }\href {https://doi.org/10.1103/PhysRevB.88.085318}
  {\bibfield  {journal} {\bibinfo  {journal} {Phys. Rev. B}\ }\textbf {\bibinfo
  {volume} {88}},\ \bibinfo {pages} {085318} (\bibinfo {year}
  {2013})}\BibitemShut {NoStop}%
\bibitem [{\citenamefont {Galvani}\ \emph {et~al.}(2016)\citenamefont
  {Galvani}, \citenamefont {Paleari}, \citenamefont {Miranda}, \citenamefont
  {Molina-S\'anchez}, \citenamefont {Wirtz}, \citenamefont {Latil},
  \citenamefont {Amara},\ and\ \citenamefont {Ducastelle}}]{Alejandro2016}%
  \BibitemOpen
  \bibfield  {author} {\bibinfo {author} {\bibfnamefont {T.}~\bibnamefont
  {Galvani}}, \bibinfo {author} {\bibfnamefont {F.}~\bibnamefont {Paleari}},
  \bibinfo {author} {\bibfnamefont {H.~P.~C.}\ \bibnamefont {Miranda}},
  \bibinfo {author} {\bibfnamefont {A.}~\bibnamefont {Molina-S\'anchez}},
  \bibinfo {author} {\bibfnamefont {L.}~\bibnamefont {Wirtz}}, \bibinfo
  {author} {\bibfnamefont {S.}~\bibnamefont {Latil}}, \bibinfo {author}
  {\bibfnamefont {H.}~\bibnamefont {Amara}},\ and\ \bibinfo {author}
  {\bibfnamefont {F.~m.~c.}\ \bibnamefont {Ducastelle}},\ }\href
  {https://doi.org/10.1103/PhysRevB.94.125303} {\bibfield  {journal} {\bibinfo
  {journal} {Phys. Rev. B}\ }\textbf {\bibinfo {volume} {94}},\ \bibinfo
  {pages} {125303} (\bibinfo {year} {2016})}\BibitemShut {NoStop}%
\bibitem [{\citenamefont {Di~Sabatino}\ \emph {et~al.}(2020)\citenamefont
  {Di~Sabatino}, \citenamefont {Berger},\ and\ \citenamefont
  {Romaniello}}]{di2020optical}%
  \BibitemOpen
  \bibfield  {author} {\bibinfo {author} {\bibfnamefont {S.}~\bibnamefont
  {Di~Sabatino}}, \bibinfo {author} {\bibfnamefont {J.}~\bibnamefont
  {Berger}},\ and\ \bibinfo {author} {\bibfnamefont {P.}~\bibnamefont
  {Romaniello}},\ }\href@noop {} {\bibfield  {journal} {\bibinfo  {journal}
  {Faraday Discussions}\ }\textbf {\bibinfo {volume} {224}},\ \bibinfo {pages}
  {467} (\bibinfo {year} {2020})}\BibitemShut {NoStop}%
\bibitem [{\citenamefont {Zhang}\ \emph {et~al.}(2018)\citenamefont {Zhang},
  \citenamefont {Shan},\ and\ \citenamefont {Xiao}}]{Zhang2018}%
  \BibitemOpen
  \bibfield  {author} {\bibinfo {author} {\bibfnamefont {X.}~\bibnamefont
  {Zhang}}, \bibinfo {author} {\bibfnamefont {W.-Y.}\ \bibnamefont {Shan}},\
  and\ \bibinfo {author} {\bibfnamefont {D.}~\bibnamefont {Xiao}},\ }\href
  {https://doi.org/10.1103/PhysRevLett.120.077401} {\bibfield  {journal}
  {\bibinfo  {journal} {Phys. Rev. Lett.}\ }\textbf {\bibinfo {volume} {120}},\
  \bibinfo {pages} {077401} (\bibinfo {year} {2018})}\BibitemShut {NoStop}%
\bibitem [{\citenamefont {Cao}\ \emph {et~al.}(2018)\citenamefont {Cao},
  \citenamefont {Wu},\ and\ \citenamefont {Louie}}]{Cao2018}%
  \BibitemOpen
  \bibfield  {author} {\bibinfo {author} {\bibfnamefont {T.}~\bibnamefont
  {Cao}}, \bibinfo {author} {\bibfnamefont {M.}~\bibnamefont {Wu}},\ and\
  \bibinfo {author} {\bibfnamefont {S.~G.}\ \bibnamefont {Louie}},\ }\href
  {https://doi.org/10.1103/PhysRevLett.120.087402} {\bibfield  {journal}
  {\bibinfo  {journal} {Phys. Rev. Lett.}\ }\textbf {\bibinfo {volume} {120}},\
  \bibinfo {pages} {087402} (\bibinfo {year} {2018})}\BibitemShut {NoStop}%
\bibitem [{\citenamefont {Taghizadeh}\ and\ \citenamefont
  {Pedersen}(2019)}]{Pedersen2019}%
  \BibitemOpen
  \bibfield  {author} {\bibinfo {author} {\bibfnamefont {A.}~\bibnamefont
  {Taghizadeh}}\ and\ \bibinfo {author} {\bibfnamefont {T.~G.}\ \bibnamefont
  {Pedersen}},\ }\href {https://doi.org/10.1103/PhysRevB.99.235433} {\bibfield
  {journal} {\bibinfo  {journal} {Phys. Rev. B}\ }\textbf {\bibinfo {volume}
  {99}},\ \bibinfo {pages} {235433} (\bibinfo {year} {2019})}\BibitemShut
  {NoStop}%
\bibitem [{\citenamefont {Pedersen}(2015)}]{Thomas2015cond}%
  \BibitemOpen
  \bibfield  {author} {\bibinfo {author} {\bibfnamefont {T.~G.}\ \bibnamefont
  {Pedersen}},\ }\href {https://doi.org/10.1103/PhysRevB.92.235432} {\bibfield
  {journal} {\bibinfo  {journal} {Phys. Rev. B}\ }\textbf {\bibinfo {volume}
  {92}},\ \bibinfo {pages} {235432} (\bibinfo {year} {2015})}\BibitemShut
  {NoStop}%
\bibitem [{\citenamefont {Rytova}(1967)}]{rytova1967}%
  \BibitemOpen
  \bibfield  {author} {\bibinfo {author} {\bibfnamefont {S.}~\bibnamefont
  {Rytova}},\ }\href@noop {} {\bibfield  {journal} {\bibinfo  {journal} {Moscow
  University Physics Bulletin}\ }\textbf {\bibinfo {volume} {22}} (\bibinfo
  {year} {1967})}\BibitemShut {NoStop}%
\bibitem [{\citenamefont {Keldysh}(1979)}]{keldysh1979coulomb}%
  \BibitemOpen
  \bibfield  {author} {\bibinfo {author} {\bibfnamefont {L.}~\bibnamefont
  {Keldysh}},\ }\href@noop {} {\bibfield  {journal} {\bibinfo  {journal} {Sov.
  J. Exp. and Theor. Phys. Lett.}\ }\textbf {\bibinfo {volume} {29}},\ \bibinfo
  {pages} {658} (\bibinfo {year} {1979})}\BibitemShut {NoStop}%
\bibitem [{\citenamefont {Li}\ and\ \citenamefont {Appelbaum}(2019)}]{Li2019}%
  \BibitemOpen
  \bibfield  {author} {\bibinfo {author} {\bibfnamefont {P.}~\bibnamefont
  {Li}}\ and\ \bibinfo {author} {\bibfnamefont {I.}~\bibnamefont {Appelbaum}},\
  }\href {https://doi.org/10.1103/PhysRevB.99.035429} {\bibfield  {journal}
  {\bibinfo  {journal} {Phys. Rev. B}\ }\textbf {\bibinfo {volume} {99}},\
  \bibinfo {pages} {035429} (\bibinfo {year} {2019})}\BibitemShut {NoStop}%
\bibitem [{\citenamefont {Tian}\ \emph {et~al.}(2019)\citenamefont {Tian},
  \citenamefont {Scullion}, \citenamefont {Hughes}, \citenamefont {Li},
  \citenamefont {Shih}, \citenamefont {Coleman}, \citenamefont {Chhowalla},\
  and\ \citenamefont {Santos}}]{tian2019electronic}%
  \BibitemOpen
  \bibfield  {author} {\bibinfo {author} {\bibfnamefont {T.}~\bibnamefont
  {Tian}}, \bibinfo {author} {\bibfnamefont {D.}~\bibnamefont {Scullion}},
  \bibinfo {author} {\bibfnamefont {D.}~\bibnamefont {Hughes}}, \bibinfo
  {author} {\bibfnamefont {L.~H.}\ \bibnamefont {Li}}, \bibinfo {author}
  {\bibfnamefont {C.-J.}\ \bibnamefont {Shih}}, \bibinfo {author}
  {\bibfnamefont {J.}~\bibnamefont {Coleman}}, \bibinfo {author} {\bibfnamefont
  {M.}~\bibnamefont {Chhowalla}},\ and\ \bibinfo {author} {\bibfnamefont
  {E.~J.}\ \bibnamefont {Santos}},\ }\href@noop {} {\bibfield  {journal}
  {\bibinfo  {journal} {Nano letters}\ }\textbf {\bibinfo {volume} {20}},\
  \bibinfo {pages} {841} (\bibinfo {year} {2019})}\BibitemShut {NoStop}%
\bibitem [{\citenamefont {Laturia}\ \emph {et~al.}(2018)\citenamefont
  {Laturia}, \citenamefont {Van~de Put},\ and\ \citenamefont
  {Vandenberghe}}]{laturia2018dielectric}%
  \BibitemOpen
  \bibfield  {author} {\bibinfo {author} {\bibfnamefont {A.}~\bibnamefont
  {Laturia}}, \bibinfo {author} {\bibfnamefont {M.~L.}\ \bibnamefont {Van~de
  Put}},\ and\ \bibinfo {author} {\bibfnamefont {W.~G.}\ \bibnamefont
  {Vandenberghe}},\ }\href@noop {} {\bibfield  {journal} {\bibinfo  {journal}
  {npj 2D Materials and Applications}\ }\textbf {\bibinfo {volume} {2}},\
  \bibinfo {pages} {1} (\bibinfo {year} {2018})}\BibitemShut {NoStop}%
\bibitem [{\citenamefont {Gon{\c{c}}alves}\ and\ \citenamefont
  {Peres}(2016)}]{gonccalves2016introduction}%
  \BibitemOpen
  \bibfield  {author} {\bibinfo {author} {\bibfnamefont {P.~A.~D.}\
  \bibnamefont {Gon{\c{c}}alves}}\ and\ \bibinfo {author} {\bibfnamefont
  {N.~M.}\ \bibnamefont {Peres}},\ }\href@noop {} {\emph {\bibinfo {title} {An
  introduction to graphene plasmonics}}}\ (\bibinfo  {publisher} {World
  Scientific},\ \bibinfo {year} {2016})\BibitemShut {NoStop}%
\bibitem [{\citenamefont {Sauer}\ and\ \citenamefont
  {Pedersen}(2021)}]{sauer2021exciton}%
  \BibitemOpen
  \bibfield  {author} {\bibinfo {author} {\bibfnamefont {M.~O.}\ \bibnamefont
  {Sauer}}\ and\ \bibinfo {author} {\bibfnamefont {T.~G.}\ \bibnamefont
  {Pedersen}},\ }\href@noop {} {\bibfield  {journal} {\bibinfo  {journal}
  {arXiv preprint arXiv:2110.14428}\ } (\bibinfo {year} {2021})}\BibitemShut
  {NoStop}%
\bibitem [{\citenamefont {Chao}\ and\ \citenamefont {Chuang}(1991)}]{Chao1991}%
  \BibitemOpen
  \bibfield  {author} {\bibinfo {author} {\bibfnamefont {C.~Y.-P.}\
  \bibnamefont {Chao}}\ and\ \bibinfo {author} {\bibfnamefont {S.~L.}\
  \bibnamefont {Chuang}},\ }\href {https://doi.org/10.1103/PhysRevB.43.6530}
  {\bibfield  {journal} {\bibinfo  {journal} {Phys. Rev. B}\ }\textbf {\bibinfo
  {volume} {43}},\ \bibinfo {pages} {6530} (\bibinfo {year}
  {1991})}\BibitemShut {NoStop}%
\bibitem [{\citenamefont {Kythe}\ and\ \citenamefont
  {Puri}(2011)}]{kythe2011computational}%
  \BibitemOpen
  \bibfield  {author} {\bibinfo {author} {\bibfnamefont {P.}~\bibnamefont
  {Kythe}}\ and\ \bibinfo {author} {\bibfnamefont {P.}~\bibnamefont {Puri}},\
  }\href@noop {} {\emph {\bibinfo {title} {Computational methods for linear
  integral equations}}}\ (\bibinfo  {publisher} {Springer Science \& Business
  Media},\ \bibinfo {year} {2011})\BibitemShut {NoStop}%
\end{thebibliography}
%

\end{document}